\documentclass[prl,twocolumn,superscriptaddress,notitlepage,floatfix,amssymb]{revtex4-1}

\usepackage[letterpaper,hmargin={1.5cm,1.5cm},vmargin={1.5cm,2.5cm}]{geometry}
\usepackage{mathtools}
\usepackage{float}
\usepackage{amsfonts}
\usepackage{enumitem}
\usepackage{graphicx}
\usepackage{bm}
\usepackage{amssymb}
\usepackage{graphicx}
\usepackage{epstopdf}
\usepackage{bmpsize}
\usepackage{epsfig}
\usepackage{hyperref}
\usepackage{float}
\usepackage{braket}
\usepackage{lipsum}
\usepackage{xcolor}
\usepackage{marginnote}
\usepackage[author=smbdy]{pdfcomment}
\usepackage{xcolor}

\begin{document}
\title{Quantum Simulation of the Bosonic Creutz Ladder with a Parametric Cavity}

\author{Jimmy S.C. Hung}
\thanks{JSCH and JHB contributed equally to this work.}
\affiliation{Institute for Quantum Computing and Department of Electrical \& Computer Engineering, University of Waterloo, Waterloo, Ontario, N2L 3G1, Canada}

\author{J.H. Busnaina}
\thanks{JSCH and JHB contributed equally to this work.}
\affiliation{Institute for Quantum Computing and Department of Electrical \& Computer Engineering, University of Waterloo, Waterloo, Ontario, N2L 3G1, Canada}

\author{C.W. Sandbo Chang}
\affiliation{Institute for Quantum Computing and Department of Electrical \& Computer Engineering, University of Waterloo, Waterloo, Ontario, N2L 3G1, Canada}

\author{A.M. Vadiraj}
\affiliation{Institute for Quantum Computing and Department of Electrical \& Computer Engineering, University of Waterloo, Waterloo, Ontario, N2L 3G1, Canada}

\author{I. Nsanzineza}
\affiliation{Institute for Quantum Computing and Department of Electrical \& Computer Engineering, University of Waterloo, Waterloo, Ontario, N2L 3G1, Canada}

\author{E. Solano}
\affiliation{IQM, Nymphenburgerstr. 86, 80636 Munich, Germany}
\affiliation{International Center of Quantum Artificial Intelligence for Science and Technology (QuArtist) \\ and Department of Physics, Shanghai University, 200444 Shanghai, China}
\affiliation{Department of Physical Chemistry, University of the Basque Country UPV/EHU, Apartado 644, 48080 Bilbao, Spain}
\affiliation{IKERBASQUE, Basque Foundation for Science, Plaza Euskadi 5, 48009 Bilbao, Spain}

\author{H. Alaeian}
\affiliation{Departments of Electrical \& Computer Engineering and Physics \& Astronomy, Purdue University, West Lafayette, IN 47907, USA}

\author{E. Rico}
\affiliation{Department of Physical Chemistry, University of the Basque Country UPV/EHU, Apartado 644, 48080 Bilbao, Spain}
\affiliation{IKERBASQUE, Basque Foundation for Science, Plaza Euskadi 5, 48009 Bilbao, Spain}

\author{C.M. Wilson}
\email{chris.wilson@uwaterloo.ca}
\affiliation{Institute for Quantum Computing and Department of Electrical \& Computer Engineering, University of Waterloo, Waterloo, Ontario, N2L 3G1, Canada}

\begin{abstract}
There has been a growing interest in realizing quantum simulators for physical systems where perturbative methods are ineffective. The scalability and flexibility of circuit quantum electrodynamics (cQED) make it a promising platform to implement various types of simulators, including lattice models of strongly-coupled field theories. Here, we use a multimode superconducting parametric cavity as a hardware-efficient analog quantum simulator, realizing a lattice in synthetic dimensions with complex hopping interactions. The coupling graph, \textit{i.e.} the realized model, can be programmed \textit{in situ}. The complex-valued hopping interaction further allows us to simulate, for instance, gauge potentials and topological models. As a demonstration, we simulate a plaquette of the bosonic Creutz ladder. We characterize the lattice with scattering measurements, reconstructing the experimental Hamiltonian and observing emerging topological features. This platform can be easily extended to larger lattices and different models involving other interactions.
\end{abstract}

\date{\today}

\maketitle


With large-scale, error-corrected quantum computers still years away, there has been considerable recent attention in analog quantum simulation (AQS)~\cite{Sorensen99,Luengo19,Blais04,Blais20,Lee05,Lanyon10,Guzik05,Mei18}. AQS is a paradigm of quantum computation where a well-controlled artificial system, \textit{e.g.} a quantum circuit, is constructed to have the same Hamiltonian as a system of interest~\cite{Islam15,Bernien17,Barreiro11,Brown16,Braumuller17,Langford17}.  The dynamics can then be explored by studying the artificial system, \textit{i.e.}, the simulator. Like analog classical computation in the 1960s, AQS is a promising path to unlock the advantages of quantum computing before large-scale digital quantum computers become feasible.

There is a particular interest in performing quantum simulations of systems that are classically intractable.  A broad class of such problems are strongly-coupled quantum field theories. These theories include fundamental models such as quantum chromodynamics (QCD), but are also our language to describe a wide array of quantum materials such as high-temperature superconductors.  Due to their strong interactions, these theories are not amenable to the standard tools of perturbation theory, making the development of simulation tools critical. While very powerful classical simulation tools exist for some of these problems~\cite{Ran20,Zohar18}, such as lattice QCD, some important problems remain intractable, for instance, due to the infamous “sign problem”~\cite{Dornheim19}. Since topological models are often affected by the sign problem, they are natural candidates for AQS~\cite{Sulejmanpasic19,Banuls20}.  

In this work, we present an \textit{in situ} programmable platform for analog quantum simulation. As a demonstration of its potential, we use it for a small-scale simulation of the bosonic Creutz ladder (BCL)~\cite{Creutz99, Alaeian19,Zurita20}. The Creutz ladder is a simple quasi-1D lattice model, but nonetheless exhibits a wide range of interesting behavior including topological and chiral states. It is historically important as one of the first models of chiral lattice fermions~\cite{Creutz99,Kaplan12}. 

The platform we use for AQS is a multimode superconducting parametric cavity. The device has several resonant modes that share a common boundary condition, which is imposed by a superconducting quantum interference device (SQUID). By modulating the shared boundary in time, we induce parametric couplings between modes, including standard ``hopping" terms~\cite{Chang18,Chang20,Lecocq17}. By selecting a set of modulation frequencies, we can create a programmable graph of connections between the modes, which then become the nodes of our lattice arrayed in synthetic dimensions. Because the couplings are created by coherent pump tones, we can control not only the magnitudes of the hopping terms, but also their relative phases. This phase control allows us to implement models with complex hopping terms, describing classical gauge fields and a variety of topological systems. We further reconstruct the realized Hamiltonian through detailed scattering measurements. We note a recent work that looked at AQS in a parametric cavity, but with a pumping scheme that lacked the addressibility and phase control demonstrated here~\cite{Lee20}. 

\begin{figure}
\center
\includegraphics[width=1\linewidth]{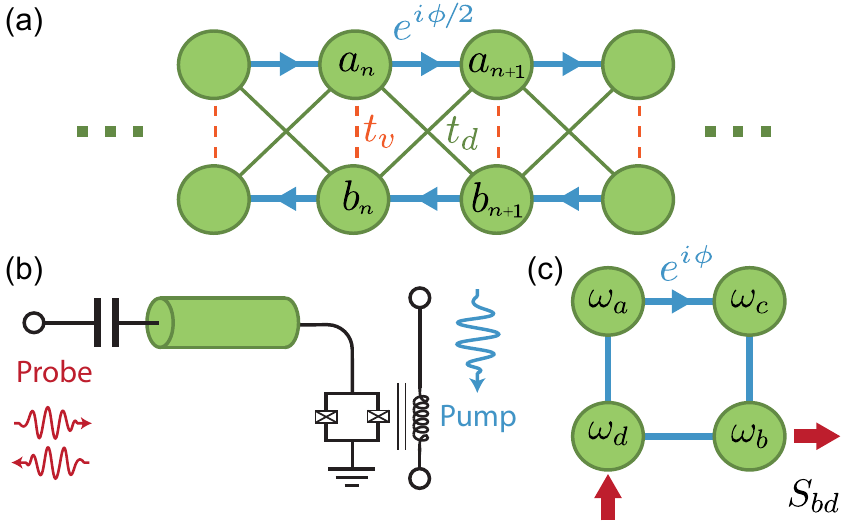}
\caption{(a) Schematic representation of the Creutz ladder. The arrows indicate the sign of the hopping phase. (b) Device cartoon. We realize interactions between cavity modes by parametrically pumping the SQUID through a flux line. The system is then probed through the input capacitor by a coherent tone. (c) Synthetic lattice. We program a four-node lattice in synthetic dimensions using four pump tones, which have a well-controlled phase. We measure the scattering matrix of the system by probing near each node frequency and measuring the output at various nodes, which are separated in frequency space.}
\label{fig:Device}
\end{figure}

The Hamiltonian of the infinite Creutz ladder, illustrated in Fig.~\ref{fig:Device}(a), is ($\hbar = 1$)
\begin{equation}
\begin{split}
\hat{\mathcal{H}}_C =-\sum_n &\Big[  t_d \left(\hat{b}_n^\dagger \hat{a}_{n+1}  + \hat{a}_n^\dagger \hat{b}_{n+1}\right) + \frac{t_v}{2} \left(\hat{b}_n^\dagger \hat{a}_n + \hat{a}_{n+1}^\dagger \hat{b}_{n+1}\right) \\
&+t_h e^{i\frac{\phi}{2}} \left(\hat{a}_{n+1}^\dagger \hat{a}_n + \hat{b}_n^\dagger \hat{b}_{n+1}\right)\Big] + {\rm H.C.},
\label{eq:CrHam}
\end{split}
\end{equation}
where $t_d,t_v, t_h$ are the diagonal, vertical, and horizontal coupling rates and $\phi/2$ is the phase of the horizontal coupling. This Hamiltonian describes the dynamics of a crossed-link fermionic ladder in a magnetic field~\cite{Creutz99}. There are a number of interesting topological features of the model. As elaborated in our previous work~\cite{Alaeian19}, at $\phi = \pi$, the Hamiltonian is time-reversal, particle-hole, and chiral symmetric. Moreover, in the so-called strong coupling limit of $t_v = 0$ and $t_d = t_h = 1$ and with open boundary conditions (finite chain), there are two chiral zero-energy modes localized at the two ends of the ladder. Here, we study the simplest building block one can use to investigate the chiral properties of the Creutz ladder. 

We can program the bosonic version of $\hat{\mathcal{H}}_C$ into our parametric cavity with the appropriate choice of pump frequencies. For ease of notation, we will now drop the $\{\hat{a}_n,\hat{b}_n\}$ notation of Eqn.~\ref{eq:CrHam} and simplify to $\{\hat{a}_n\}$ with the connectivity of the lattice now encoded in a coupling tensor $g_{nm}$. 

To probe the system, we must couple it to our measurement line, which we model with the coupling Hamiltonian
 \begin{equation}~\label{eq:coherent drive}
 \hat{\mathcal{H}}_P = i \sum_n \sqrt{\kappa_n^{\textrm{ext}}}   \left( \hat{a}_{i,n}  - \hat{a}_{i,n}^\dagger \right) \left( \hat{a}_n  + \hat{a}_n^\dagger \right),
\end{equation}
with $\hat{a}_{i,n}$ describing the annihilation operator of the $n^{th}$ input mode with the external coupling rate $\kappa_{n}^{\textrm{ext}}$.
To treat the dynamics of our driven, dissipative system, we use the following Lindblad master equation~\cite{Lindblad76,Gardiner04}
\begin{equation*}~\label{eq:Lindblad ME}
\dot{\hat{\rho}} = -i \left[\hat{\mathcal{H}}_C + \hat{\mathcal{H}}_P,\hat{\rho}\right] + \sum_n \kappa_n \left(\hat{a}_n \hat{\rho} \hat{a}_n^\dagger - \frac{1}{2}\{\hat{a}_n^\dagger \hat{a}_n, \hat{\rho}\}\right),
 \end{equation*}
  where $\hat{\rho}$ is the reduced density matrix of the plaquette and $\kappa_n = \kappa_{n}^{\textrm{ext}} + \kappa_{n}^{\textrm{int}}$ is the total photon decay rate including the internal loss rate $\kappa_{n}^{\textrm{int}}$~\footnote{Here we are mainly interested in the dynamics of cavity modes so the transmission-line bath can be integrated out in the typical Lindblad form. The material loss inside the cavity leads to additional decay which we could either be included as a non-Hermitian term $\kappa_{n}^{\textrm{int}}$ or directly incorporated in the Lindblad form. If the latter, their noise input contribution and corresponding input/output formalism must be treated differently. While the coupling to the transmission line leads to input noise terms, the internal loss noise contribution should be governed by the fluctuation-dissipation theorem at equilibrium temperature.}.
 
The Heisenberg-Langevin equations of motion for the mode operators follow directly as
\begin{equation}~\label{eq:HL EoM}
\dot{\hat{a}}_n = i \left(\Delta_n + i\frac{\kappa_n}{2} \right)\hat{a}_n + i\sum_{m \ne n} \frac{g_{nm}}{2} \hat{a}_m + \sqrt{\kappa_{n}^{\textrm{ext}}} \hat{a}_{i,n},
\end{equation}
where $\Delta_n = \omega_n^s - \omega_n$ with $\omega_n^s$ the probe frequency of the $n^{th}$ mode. Using the input-output formalism, the output modes which we detect are then defined as $\hat{a}_{o,n} = \sqrt{\kappa_{n}^{\textrm{ext}}} \hat{a}_n - \hat{a}_{i,n}$. Finally, to find the scattering matrix, we solve for the steady-state solutions of Eqn.~\ref{eq:HL EoM}, \textit{i.e.}, assuming $\dot{\hat{a}}_n =0$, and define $S_{nm} = \braket{\hat{a}_{o,n}}/\braket{\hat{a}_{i,m}}$. We note that the same scattering equations can be derived from an effective non-Hermitian Hamiltonian~\cite{Lecocq17,SupNote}.

The parametric cavity is a quarter wavelength coplanar waveguide resonator terminated by a SQUID at one end and capacitively overcoupled (Q $\approx$ 7000) to a 50 $\Omega$ transmission line at the other end \cite{Chang18,Chang20,Sandberg08,Wilson10,Simoen15}. The fundamental mode of the cavity is designed to be around 1~GHz which results in five accessible modes within our measurement bandwidth of 4-12~GHz. Impedance engineering of the cavity makes the mode spacing nondegenerate, enabling frequency selective activation of parametric interactions between specific modes \cite{Bajjani11,Chang18}. We can activate a variety of parametric processes by modulating the boundary condition of the cavity using a microwave pump, which is coupled to the SQUID~\cite{Chang18,Chang20,Lecocq17}. For this work, the hopping terms are activated by pumping at the difference of two mode frequencies. 

%
\begin{figure*}
\center
\includegraphics[width=1\linewidth]{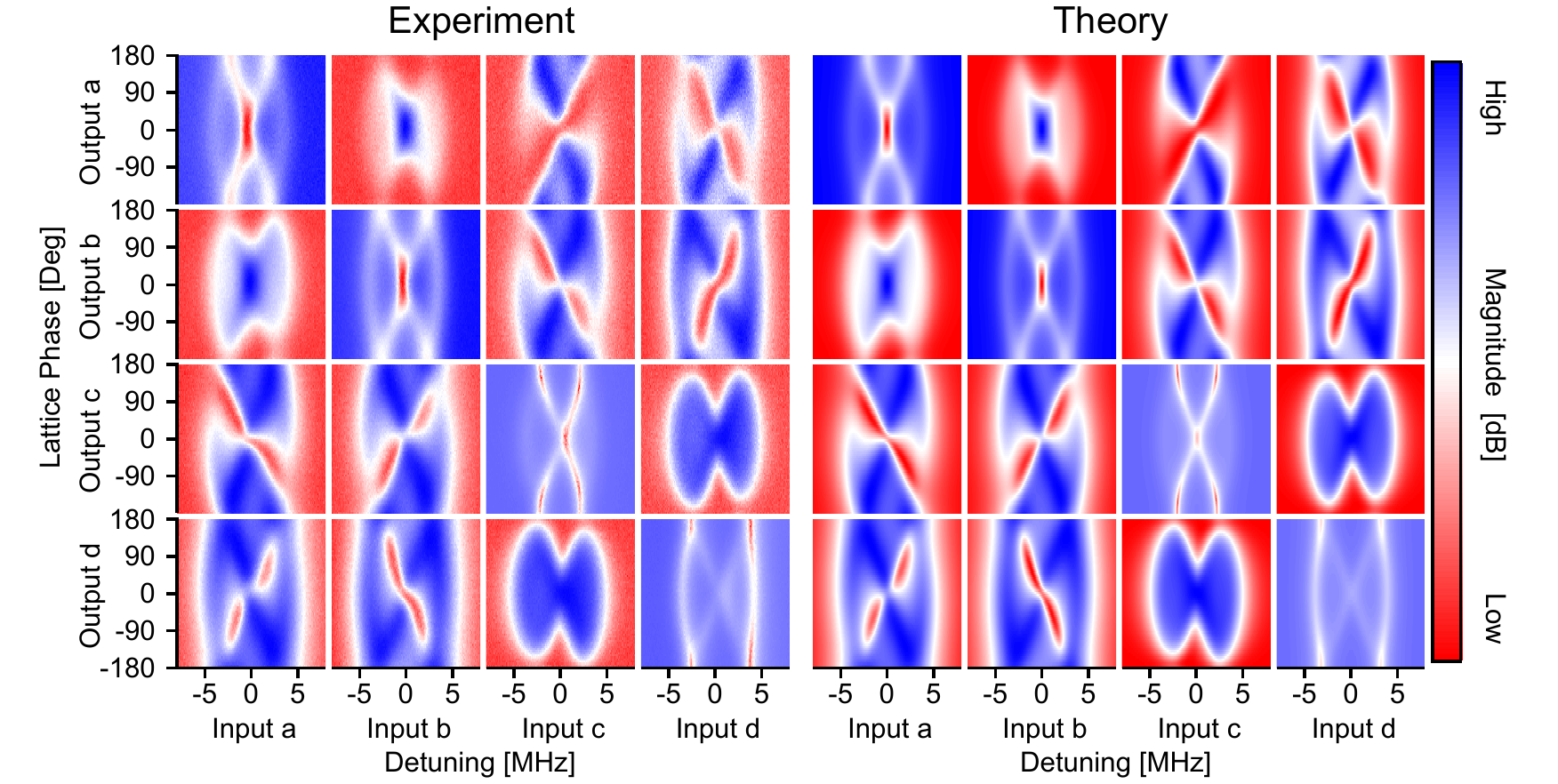}
\caption{The scattering matrix. The magnitude of the experimental (left) and theoretical (right) scattering matrices as a function of phase and frequency. The frequency axes give the detunings from the uncoupled mode frequencies. The diagonal reflection coefficients, $\{S_{nn}\}$, provide the spectrum the lattice eigenmodes. The off-diagonal elements, $\{S_{mn}\}$, are the magnitude of frequency-converting transport between the nodes. The $\{S_{mn}\}$ allows us to characterize the ``spatial" support of each eigenmode over the lattice in the synthetic dimensions (see text). We see clearly that the transport is nonreciprocal with $\{S_{mn}\}$ and $\{S_{nm}\}$ often being complements of each other.}
\label{fig:scattering}
\end{figure*}
We program a four-node plaquette by pumping the SQUID with four coherent tones at the appropriate difference frequencies, $\omega_{nm}^p$, as seen in Fig.~\ref{fig:Device}~(c).  The choice of $\omega_{nm}^p$ determine which $g_{nm}$ are nonzero, programming the connection graph of the lattice. As described below, we can associate this small lattice with various parts of a larger Creutz ladder. The mode and pump frequencies are listed in Table~\ref{tab:fitting}, where we use the specific mode labels $n \in$ \{a,b,c,d\}. We generate the pump tones using microwave generators phase-locked using 1~GHz references, which provides superior phase coherence. 

To characterize the lattice, we use a vector network analyzer (VNA) to probe the system through its input capacitor. To measure the reflection coefficient, $S_{nn}(\Delta_n)$, of node $n$, we both probe and detect around that node's frequency, $\omega_n$. When the lattice is activated, the single resonance observed at each uncoupled mode frequency is split into a number of resonances. We can interpret this set of resonances as the spectrum of the eigenmodes that exist on the lattice. Each element is centered on the uncoupled mode frequency and the frequency offset of the coupled eigenmodes can be viewed as the energy of the mode in the common rotating frame of the pumps. We can infer the mode coupling strengths, the $g_{nm}$ of Eqn.~\ref{eq:HL EoM}, as a function of pump power from the set of spectra $\{S_{nn}\}$. For the simple case of two coupled modes, the frequency splitting of the eigenmodes directly gives the coupling strength. The situation is more complicated with more than two modes, but the basic intuition is similar. 

In setting the coupling strengths for the lattice links, we normalize the coupling strengths to the geometric mean of the mode linewidths $\kappa_m$ and $\kappa_n$, defining $\beta_{nm} = g_{nm}/2\sqrt{\kappa_m \kappa_n}$. Here, we chose the $\beta_{nm}$ to be roughly equal and in the strong-coupling limit. We use strong coupling here to mean that the eigenmodes of the system are resolved in frequency, as seen in Fig.~\ref{fig:scattering}.
Since different lattice nodes exist along synthetic dimensions in frequency space, measuring the off-diagonal scattering coefficients $S_{mn}$, which characterize transport between nodes, requires a frequency-conversion measurement, where the probe and detection frequency are different. We can distinguish $S_{mn}$ and $S_{nm}$ by swapping the probe and detection frequencies, allowing us to see nonreciprocal features in the transport.

We simulate the effect of applying an external magnetic field to the lattice by making the $\{g_{nm}\}$ complex. The phase of the hopping term represents the phase acquired by an excitation moving along the link in the presence of the field. For our simple 4-node plaquette, only the total phase around the loop matters.  As such, we choose, without loss of generality, to sweep the phase, $\phi$, of the hopping term between modes a and c. Formally, moving the phases between links can be seen as a gauge transformation.

Figure \ref{fig:scattering} shows the measured 4x4 scattering matrix. Each element $S_{nm}$ is measured as a function of $\phi$ and $\Delta_n$.  We clearly see nontrivial behavior as $\phi$ is varied, with a series of degeneracy arising and disappearing. The off-diagonal elements $\{S_{mn}\}$ show the magnitude of frequency-converting transport from node $n$ to node $m$.  The frequency differences are set by the pump frequencies, $\omega_{nm}^p$ (see Table~\ref{tab:fitting}). These transport measurements allow us to recover ``spatial" information about the support of the eigenmodes over the synthetic lattice. Being in the strong-coupling limit, we can excite a specific eigenmode at a well-defined detuning, $\Delta_n$. As the eigenmodes are ``spatially" distributed along the synthetic dimensions, the excitation hops between the nodes and eventually leaks out of the cavity at another node, where it is then detected at the converted frequency. 

We performed detailed fitting of the data in Fig.~\ref{fig:scattering}. To extract the model parameters, we fit all of the scattering elements simultaneously at several phases~\cite{SupNote}.  In total, the fit was done to 64 VNA traces simultaneously, so, while the number of parameters is substantial, there is a large amount of data to constrain the fit. Figure \ref{fig:scattering} also shows the fit scattering matrix.  Table~\ref{tab:fitting} shows the extracted parameters. We find that the quality of the fit is remarkable given the complexity of the data.

We observe a number of interesting features in the scattering matrix. First, we observe clear nonreciprocity in the transport, for instance, noticing that $S_{bc}$ and $S_{cb}$ are effectively complements of each other. The definition of reciprocity is that $S_{ij} = S_{ji}$ which is clearly broken here. We will not emphasize it here, but this can be connected to the fictitious magnetic flux breaking time-reversal symmetry. 

\setlength{\tabcolsep}{6pt}
\begin{table}[H]
    \centering
    \begin{tabular}{c | c c c c}
         \toprule 
         Mode & a & b & c & d \\
         \hline 
          $\omega_n/ 2\pi$ [GHz] & 4.1589 &6.0992 &7.4726 & 9.4806 \\
          $\kappa_n / 2\pi$ [MHz] & 1.0147 & 1.6533 & 2.9161 & 4.5858 \\
         $\kappa_n^{ext} / 2\pi$ [MHz] & 0.690 & 1.223 & 2.566 & 3.485 \\
         \hline \hline
         Coupling & ac & ad & bc & bd \\
         \hline
         $\omega_{nm}^p/ 2\pi$ [GHz] & 3.3136 & 5.3223 & 1.3733 & 3.382 \\
         $|g_{nm}| / 2\pi$ [MHz] & 2.9077 &  3.7107 & 3.4973 & 5.6458 \\
         $\beta_{nm}$ & 0.8452  & 0.8601  & 0.7964 & 1.0252 \\
                  \hline \hline
    \end{tabular}
    \caption{Extracted device and lattice parameters.}
    \label{tab:fitting}
\end{table}

\begin{figure}
\center
\includegraphics[width=1\linewidth]{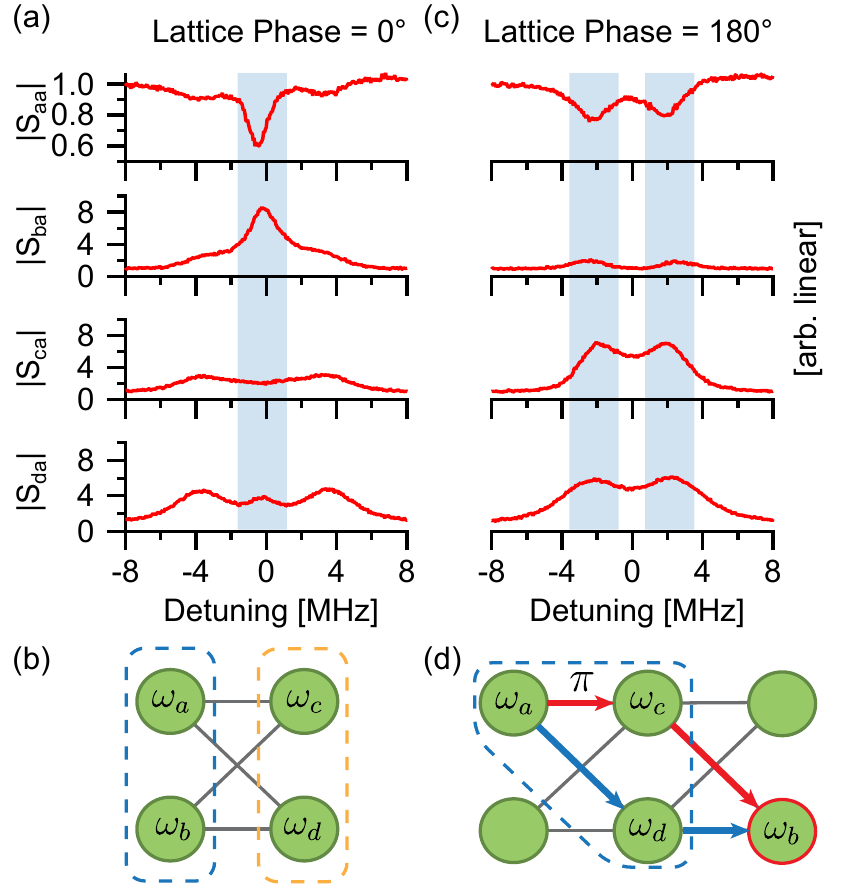}
\caption{Emerging topological features. (a) $\phi=0$ line cuts of the scattering parameters in Fig.~\ref{fig:scattering} when probing at node a. The vertical axes are normalized to the background. The measurements indicate the existence of an eigenmode at zero energy with significant support only in the nonadjacent nodes a and b. We infer this from the relatively high transmission amplitude from a to b. We also observe a 2nd zero mode localized on sites c and d. (b) Twisted plaquette. We expect topological features of the Creutz ladder to appear at $\phi = \pi$ and not $\phi = 0$.  However, we note that if we twist the plaquette as indicated, $\phi = 0$ regardless of the external flux. After twisting, the zero modes now appear at the two ends of the plaquette. These states are reminiscent of the predicted zero-mode end states \cite{Creutz99}. (c) $\phi=\pi$ line cuts when probing from node a. The scales of the vertical axes are the same as panel (a). The measurements indicate the existence of two pairs of degenerate eigenmodes, one at positive and one at negative detuning, that have support on all but one of the nodes. One of these four eigenmodes exists on each corner.  (d) Caging. We can associate the corner eigenmodes with the soliton states in the Creutz ladder by identifying the lattice as the indicated trapezoidal path. Due to Aharonov-Bohm caging, an excitation at, \textit{e.g.}, node a cannot propagate to node b.}
\label{fig:States}
\end{figure}

We can also identify interesting eigenmodes that we associate with emerging topological features of the Creutz ladder~\footnote{A common way to identify topological states is through an invariant such as the Chern number. While the value of these invariants can be inferred from the fully reconstructed Hamiltonian we have extracted, our theoretical investigations suggest that these invariants do not carry meaningful information on a single single plaquette.}. At $\phi = \pi$, the Creutz ladder predicts that the bulk states collapse in a pair of flat bands at equal but opposite energies. A flat band implies that the bulk states are localized, as the group velocity goes to zero. Creutz referred to the associated states as ``solitons" and identified the localization as arising from interference between alternate paths on the lattice (see Fig.~\ref{fig:States}~(d)), a phenomenon often referred to as Aharnov-Bohm caging in recent literature. With open boundary conditions, theory also predicts the existence of a pair zero-energy states localized to the ends of the ladder. The connection between the observed eigenmodes and these topological states is discussed in detail in Fig.~\ref{fig:States}.

In this letter, we have introduced a platform for programmable analog quantum simulation of topological lattice models. The platform is hardware efficient, creating the lattice in synthetic dimensions within a single parametric cavity. We have demonstrated the potential of the platform by performing small-scale simulations of a paradigmatic topological model, showing that we can reconstruct the experimentally realized Hamiltonian through scattering measurements. The obvious next step is to increase the size of the lattice. We can do this in a straightforward manner by increasing the physical length of the cavity, increasing the density of the cavity modes in frequency. Multiple cavities can also be coupled together to further scale the simulations. Efforts to implement these improvements are underway, as well as experiments on other lattice models programmed into the same device. 

\section*{Acknowledgment}
The authors wish to thank B. Plourde, J.J. Nelson and M. Hutchings at Syracuse University for invaluable help in junction fabrication. H.A. acknowledges the financial supports from the Eliteprogram award of Baden-W\"urttemberg Stiftung and Purdue university Start-up grant. E.R. thanks the QuantERA project QTFLAG, and support of the Basque Government grant IT986-16. E.S acknowledges support from QMiCS (820505), OpenSuperQ (820363) of the EU Flagship on Quantum Technologies, Spanish MINECO/FEDER FIS2015-69983-P, Basque Government IT986-16, EU FET Open Grants Quromorphic, EPIQUS, and Shanghai STCSM (Grant No. 2019SHZDZX01-ZX04). CMW, JSCH, JB, CWSC, AMV and IN acknowledge the Canada First Research Excellence Fund (CFREF), NSERC of Canada, the Canadian Foundation for Innovation, the Ontario Ministry of Research and Innovation, and Industry Canada for financial support. 

\bibliography{main,Notes}
\end{document}


\title{Supplemental material}

\author{Jimmy S.C. Hung}
\thanks{JSCH and JHB contributed equally to this work.}
\affiliation{Institute for Quantum Computing and Department of Electrical \& Computer Engineering, University of Waterloo, Waterloo, Ontario, N2L 3G1, Canada}

\author{J.H. Busnaina}
\thanks{JSCH and JHB contributed equally to this work.}
\affiliation{Institute for Quantum Computing and Department of Electrical \& Computer Engineering, University of Waterloo, Waterloo, Ontario, N2L 3G1, Canada}

\author{C.W. Sandbo Chang}
\affiliation{Institute for Quantum Computing and Department of Electrical \& Computer Engineering, University of Waterloo, Waterloo, Ontario, N2L 3G1, Canada}

\author{A.M. Vadiraj}
\affiliation{Institute for Quantum Computing and Department of Electrical \& Computer Engineering, University of Waterloo, Waterloo, Ontario, N2L 3G1, Canada}

\author{I. Nsanzineza}
\affiliation{Institute for Quantum Computing and Department of Electrical \& Computer Engineering, University of Waterloo, Waterloo, Ontario, N2L 3G1, Canada}

\author{E. Solano}
\affiliation{IQM, Nymphenburgerstr. 86, 80636 Munich, Germany}
\affiliation{International Center of Quantum Artificial Intelligence for Science and Technology (QuArtist) and Department of Physics, Shanghai University, 200444 Shanghai, China}
\affiliation{Department of Physical Chemistry, University of the Basque Country UPV/EHU, Apartado 644, 48080 Bilbao, Spain}
\affiliation{IKERBASQUE, Basque Foundation for Science, Plaza Euskadi 5, 48009 Bilbao, Spain}

\author{H. Alaeian}
\affiliation{Departments of Electrical \& Computer Engineering and Physics \& Astronomy, Purdue University, West Lafayette, IN 47907, USA}

\author{E. Rico}
\affiliation{Department of Physical Chemistry, University of the Basque Country UPV/EHU, Apartado 644, 48080 Bilbao, Spain}
\affiliation{IKERBASQUE, Basque Foundation for Science, Plaza Euskadi 5, 48009 Bilbao, Spain}

\author{C.M. Wilson}
\email{chris.wilson@uwaterloo.ca}
\affiliation{Institute for Quantum Computing and Department of Electrical \& Computer Engineering, University of Waterloo, Waterloo, Ontario, N2L 3G1, Canada}

\date{\today}

\maketitle

\section{Fitting Procedure}

After making the scattering matrix measurements, we implement basic data preprocessing steps, described in more detail below. All steps described are applied to the data in linear amplitude units, converted from logarithmic units (dB) according to the relation $S_{nm, \textrm{Lin}} = 10^{(S_{nm,\textrm{Log}}/20)}$.

Both the reflection and frequency-converting transmission measurements are subject to frequency dependent gain and loss. Our first step in eliminating these effects is to measure and correct for the background, while all pumps are off. For the reflection measurements, we detune the cavity modes out of the VNA measurement bandwidth and measure the background reflection. We then divide the cavity's reflection data by the recorded background trace.  The procedure for the transmission data is similar, except that we note that the background transmission measured is, in fact, the noise floor of the measurement system, as there is negligible frequency conversion of the probe signal if the pumps are not activated. After this background normalization, we often find that there is still a background slope in the reflection data. We remove this by fitting a line to the two ends of each reflection trace and then divided each trace by the fitted line.

With the fine-scale frequency dependence now removed, we still need to account for the large-scale (between mode) frequency dependence of the loss and gain. We do this by introducing 12 scale parameters, $C_{nm}$, one for each of the transmission coefficients. (After the background subtraction, the diagonal $C_{nn}=1$ by definition.)
Next, we incorporate the noise floor of the VNA in the transport measurement fitting by recognizing that the power of transported signal and the noise power add, giving
\begin{equation}~\label{eq:scaling}
    |S_{nm,\textrm{fit}}| = \sqrt{ \left(C_{nm} |S_{nm,{\rm Lin}}|\right)^2 + 1},
\end{equation}
recalling that the noise-floor power level has been normalized to one.

The pump tones acquire unknown phase shifts as they travel along the input lines. Since our current system has only one loop, all of these phase shifts add up to produce a constant offset in the loop phase, $\phi$. Nevertheless, based on our understanding of the system and the observed periodicity, we can digitally offset the phase of the measured data. This offset phase, $\phi_{\rm off}$, becomes a fitting parameter. In the data displayed in Fig.~2 of the main text, the extracted $\phi_{\rm off}$ has been subtracted.

The fitting was done on 2D data. The 1\textsuperscript{st} dimension is the frequency and the 2\textsuperscript{nd} is the loop phase, $\phi$. We use four different values of $\phi$: 0, $\pi/4$, $\pi/2$ and $\pi$. The phase differences between measurements are known since we control them with the microwave generators that produce the pump tones. Apart from the pump frequencies which are fixed, a total of 29 parameters are adjusted for fitting: the lattice parameters $\omega_n$, $\kappa_n$, $\eta_n$, and $\beta_{nm}$ as well as the $\phi_{\rm off}$ and the 12 scaling factors $C_{nm}$. Here we define the coupling efficiency as $\eta_n = \kappa^{ext}_n/\kappa_n $ and the normalized coupling strength $\beta_{nm} = g_{nm}/2\sqrt{\kappa_n \kappa_m}$.  The full set of extracted parameters is listed in Tables I and II.

The magnitude of 64 traces that include the 16 scattering matrix elements at four loop phase values are fit simultaneously in one global fitting routine. Note that both the upper and lower part of the scattering matrix must be fit because the transmission is nonreciprocal. The low-level fitting routine is the builtin curve fitting operation of the "Igor Pro" software package.

We use the scattering matrix in Eqn.~\ref{eq:ScatMatrix} formulated below to create the single output fitting function. The inputs of the fitting function are the probe frequency and the model parameters mentioned above. In addition, it also includes the predefined values of the pump tone frequencies and loop phases. Given those parameters, the function constructs the coupling matrix $\bold{M}$ and the coupling efficiency matrix $\bold{H}$  and then calculates the scattering matrix of the model for a defined frequency and loop phase. After that, the value of the desired scattering element, whose indices are passed as additional parameters, is extracted and rescaled based on Eqn.~\ref{eq:scaling} before it is returned as the function output.

The global fitting procedure cascades the traces of the scattering elements at the designated loop phases to create a single trace to be fit. The global fitting function is then a piece-wise function that includes the local values for loop phase and the scattering element indices as inputs. The rest of the model parameters are global.

To generate initial guesses for the fitting routine, we perform a set of measurements on pairwise coupled modes, that is, with only one coupling pump on at a time. We fit simpler models of pairwise coupled modes to this set of data to extract initial estimates of $\omega_n$, $\kappa_n$, $\eta_n$ and $\beta_{nm}$. The extracted parameters of these pairwise fits are shown in Table~\ref{tab:parameters}. 

We do the global fitting in two steps. In the first step, we hold the parameters $\omega_n$, $\kappa_n$ and $\eta_n$ constant at their initial values, and fit $\phi_{\rm off}$ along with the $\beta_n$ and $C_{nm}$.
We perform the second and final step by freeing the remaining parameters including $\omega_n$, $\kappa_n$ and $\eta_n$. We note that, comparing Table~\ref{tab:fittingParams} and Table~\ref{tab:parameters}, there are small differences between the full-lattice parameters and the pairwise parameters.  These small changes are consistent with what is expected due to the strong parametric pumping.

\setlength{\tabcolsep}{6pt}
\begin{table}[H]
    \centering
    \begin{tabular}{c | c c c c}
         \toprule 
         Mode & a & b & c & d \\
         \hline 
         $\omega_n/ 2\pi$ [GHz] & 4.1589 &6.0992 &7.4726 & 9.4806 \\
         $\sigma$ [KHz]  &  2.5  & 3.1 & 3.8 & 5.5 \\
         \hline \hline
         $\kappa_n / 2\pi$ [MHz] & 1.0147 & 1.6533 & 2.9161 & 4.5858 \\
         $\sigma$ [KHz] & 5.8& 10.7& 7.7& 15\\
         \hline \hline
         $\eta_n$ &  0.68 & 0.74  & 0.88 & 0.76  \\
         $\sigma$ & $11.7 \times 10^{-3}$  & $8 \times 10^{-3}$ & $6.3 \times 10^{-3}$ & $4.3 \times 10^{-3}$ \\
         \hline \hline
         Coupling & ac & ad & bc & bd \\
         \hline
         $\beta_{nm}$ & 0.8452  & 0.8601  & 0.7964 & 1.0252 \\
         $\sigma$ & $2.87 \times 10^{-3}$ &  $2.3 \times 10^{-3}$   &  $2.28 \times 10^{-3}$ &  $3.29 \times 10^{-3}$ \\
                  \hline \hline
    \end{tabular}
    \caption{The extracted lattice parameters and their errors, $\sigma$. See the text for definitions of the parameters.}
    \label{tab:fittingParams}
\end{table}

\setlength{\tabcolsep}{6pt}
\begin{table}[H]
    \centering
    \begin{tabular}{c | c c c c}
         \toprule 
         Mode & a & b & c & d \\
         \hline 
        a   &  \textendash &	$19.1 \pm0.36 $ &	$20.1 \pm0.41 $&	$20.1  \pm 0.36$\\
        b   & $5.8 \pm0.11 $& \textendash	&$9.5 \pm 0.12$&	$10.3 \pm0.1 $\\
        c   & $8.4 \pm 0.17$&	$12.8 \pm0.17 $&	\textendash &	$14.3 \pm0.18 $\\
        d   &  $4.3 \pm 0.09 $&	$7.2 \pm0.07 $&	$7.2 \pm0.09 $&	\textendash\\
         \hline \hline
    \end{tabular}
    \caption{The extracted scaling parameters, $C_{nm}$ and their errors.}
    \label{tab:fittingscalingParams}
\end{table}

\setlength{\tabcolsep}{6pt}
\begin{table}[H]
    \centering
    \begin{tabular}{c | c c c c}
         \toprule 
         Mode & a & b & c & d \\
         \hline 
         $\omega_n /2\pi$ [GHz] & 4.1578 & 6.0979 & 7.4719 & 9.4802 \\
         $\kappa_n / 2\pi$ [MHz] & 1.0745 & 1.6298 & 2.8179 & 4.1049 \\
         $\eta_n$ & 0.46 & 0.62 & 0.86 & 0.74 \\
         \hline \hline
         Coupling & ac & ad & bc & bd \\
         \hline
         $\omega_{nm}^p$ [GHz] & 3.3136 & 5.3223 & 1.3733 & 3.382 \\
         $g_{nm} / 2\pi$ [MHz] & 2.9795 & 3.1395 & 3.6815 & 4.6560 \\
         $\beta_{nm}$ & 0.8561 & 0.7474 & 0.8589 & 0.9000 \\
         \hline \hline
    \end{tabular}
    \caption{Extracted uncoupled and pairwise coupling parameters, used as initial guesses for the fitting routine for the full lattice. We see that there are small, $\sim 1$~MHz, frequency shifts between the pair-wise frequencies and the full lattice frequencies in Table\ref{tab:fittingParams}. These are consistent with small, pump-induced shifts of the cavity frequency.}
    \label{tab:parameters}
\end{table}

\section{Theory}

\subsection{Discrete Symmetries}

We are considering a noninteracting bosonic system described by a second quantized Hamiltonian $\hat{H}=\hat{b}^\dagger_i H^{ij} \hat{b}_j$, where the matrix $H^{ij}$ is the so called \emph{first-quantized} Hamiltonian. The  characterizing discrete symmetries of such system are 

\emph{Time-reversal Symmetry} (TR) - $\hat{T}$ is an anti-unitary operator. The system is TR invariant if $\hat{T} \hat{H} \hat{T}^{-1}=\hat{H}$. In the noninteracting case, there is a unitary $U_T$, such that $U^\dagger_T H^* U_T=H$ which in the momentum space can be written as $T H_k T^{-1} = H_{-k}$.

\emph{Charge-conjugation Symmetry} (C) - $\hat{C}$ is a unitary operator, that preserves the canonical commutation relations and fulfills $\hat{C} \hat{H} \hat{C}^{-1}=\hat{H}$ then the system is C invariant. In the noninteracting case, if the system is C invariant, there is a unitary $U_C$, such that $U^\dagger_C H^T U_C=-H$ or in the momentum space in a translation invariant model, $C H_k C^{-1} = -H_{-k}$.

\emph{Chiral  Symmetry} (S) - $\hat{S}$ is an anti-unitary operator which is the combination of the previous two symmetries, i.e. $\hat{S}=\hat{T}\hat{C}$. It preserves the canonical commutation relations hence, an S invariant system  fulfills $\hat{S} \hat{H} \hat{S}^{-1}=\hat{H}$. In the noninteracting case, if the system is S invariant, there is a unitary $U_S=U_T U_C$, such that $U^\dagger_S H^T U_S=-H$ or in the momentum space in a translation invariant model, $S H_k S^{-1} = -H_{k}$.

\subsection{Bosonic Creutz Ladder: A Summary}
Following the works by Resta and Vanderbilt~\cite{Resta07}, it is possible to obtain the topological properties of a material through the measurement of spatial quantities such as the polarization.
To get an insight about these measurements in our model, we will discuss the strong-coupling limit of the Creutz ladder defined by $t_v=0$, $t_d = t_h = 1$. At the special point, $\phi=\pi$, the Hamiltonian has the following form
\begin{equation}
\hat{\mathcal{H}}_C = -\sum_n (\hat{b}_n^\dagger \hat{a}_{n+1}  + \hat{a}_n^\dagger \hat{b}_{n+1}) 
- i \sum_n (\hat{a}_{n+1}^\dagger \hat{a}_n + \hat{b}_n^\dagger \hat{b}_{n+1}) + h.c.
\label{eq:Ham}
\end{equation}

As elaborated in our previous work~\cite{Alaeian19}, the Hamiltonian in the momentum space ($\hat{h}_k$) satisfies the following relations
\begin{equation}
    \hat{\sigma}_x \hat{h}_k \hat{\sigma}_x = \hat{h}_{-k},~
    \hat{\sigma}_z \hat{h}_k \hat{\sigma}_z = -\hat{h}_{-k} , ~
    \hat{\sigma}_y \hat{h}_k \hat{\sigma}_y = -\hat{h}_k,
    \label{eq:syms}
\end{equation}
where $\hat{\sigma}_{x,y,z}$ are the Pauli matrices. In another words, in this particular configuration the Hamiltonian is time-reversal, particle-hole, and chiral symmetric. 


The Hamiltonian can be rewritten as 
\begin{equation}
\hat{\mathcal{H}}_C= \sum_n \begin{pmatrix} \hat{\eta}^\dagger_{+,n+1/2}, & \hat{\eta}^\dagger_{-,n+1/2} \end{pmatrix} \begin{pmatrix} 2 & 0 \\ 0 & -2 \end{pmatrix} \begin{pmatrix} \hat{\eta}_{+,n+1/2} \\ \hat{\eta}_{-,n+1/2} \end{pmatrix},
\end{equation}
as can be seen, the Hamiltonian has two flat bands at $\mathcal{E} = \pm 2$ with annihilation operators of the maximally localized Wannier functions as following
\begin{equation*}
\begin{pmatrix} \hat{\eta}_{+,n+1/2} \\ \hat{\eta}_{-,n+1/2} \end{pmatrix}= \frac{1}{2} \begin{pmatrix} e^{-\frac{i\pi}{4}} \left(\hat{a}_{n+1} - \hat{b}_n \right) + e^{\frac{i\pi}{4}} \left(\hat{a}_{n} - \hat{b}_{n+1} \right)  \\ -e^{\frac{i\pi}{4}} \left(\hat{a}_{n+1} + \hat{b}_n \right) - e^{-\frac{i\pi}{4}} \left(\hat{a}_{n} + \hat{b}_{n+1} \right) \end{pmatrix}.
\end{equation*}
In the single particle picture, if we measure the position operator $\hat{m}=\sum_n n \left(\hat{a}^\dagger_n \hat{a}_n + \hat{b}^\dagger_n \hat{b}_n\right)$, which is equivalent to the polarization, for the Wannier eigenstates $|\eta_{-,n+1/2} \rangle= \eta^\dagger_{-,n+1/2} |0\rangle= \frac{1}{2} \left[ -e^{-\frac{i\pi}{4}} \left( |1_{a_{n+1}} \rangle +| 1_{b_{n}} \rangle \right) - e^{\frac{i\pi}{4}} \left( |1_{a_{n}} \rangle +| 1_{b_{n+1}} \rangle \right) \right]$
\begin{equation}
\langle \eta_{-,n+1/2} | \hat{m} |\eta_{-,n+1/2} \rangle = n+\frac{1}{2}
\end{equation}
In fact, it can be proven that $\langle \eta_{-,n+1/2} | \hat{m} |\eta_{-,n+1/2} \rangle = \frac{\phi_{Berry}}{2\pi}$

\subsection{Winding Number Measurement in a Transport Experiment}
In principle, we can extract the value of the winding number in a single-plaquette setup in a transport experiment. The Hamiltonian of Eqn.~\ref{eq:Ham} for a single plaquette is given by:
\begin{equation}
\mathcal{\hat{H}}=-\left(\hat{b}_1^\dagger \hat{a}_2 + \hat{a}_1^\dagger \hat{b}_2 + i \hat{a}_1^\dagger \hat{a}_2 + i \hat{b}_2^\dagger \hat{b}_1 + h.c. \right) =2 \left( \hat{\eta}^\dagger_+ \hat{\eta}_+ - \hat{\eta}^\dagger_- \hat{\eta}_- \right)
\end{equation}
with $\hat{\eta}_{\pm} =\frac{1}{2} \left[e^{\mp i \pi /4} \left( \hat{a}_2 \mp \hat{b}_1 \right) + e^{\pm i \pi /4} \left(\hat{a}_1 \mp \hat{b}_2 \right) \right]$ and the zero-modes $\hat{\eta}_{0,L} =\frac{1}{\sqrt{2}} \left(e^{i\pi/4} \hat{a}_1 + e^{-i\pi /4} \hat{b}_1\right)$ and $\hat{\eta}_{0,R}=\frac{1}{\sqrt{2}} \left(e^{-i\pi/4} \hat{a}_2 + e^{i\pi /4} \hat{b}_2\right)$. The time evolution of the Wannier eigenstates is given by $|+\left(t\right)\rangle = e^{-i2t}|+\rangle$, $|-\left(t\right)\rangle = e^{i2t}|-\rangle$, $|L\left(t\right)\rangle =|L\rangle$, and $|R\left(t\right)\rangle =|R\rangle$.

Consider now exciting the state $|\chi\rangle = \frac{e^{-i\pi /4}}{\sqrt{2}} \left(e^{i \pi /4}  |a_1 \rangle + e^{-i \pi /4} |b_1 \rangle \right)$, which is a superposition of $|+\rangle$ and $|-\rangle$ that lives on the left edge but is orthogonal to $|L\rangle$.  The expectation value of the position (polarization) of the excitation is given by $\langle \chi \left( t \right)| \hat{m} |\chi \left( t \right) \rangle = \frac{3-\cos{\left( 4t\right)}}{2}$, with an average value in time of $3/2$, which can be related to the winding number $n+1/2 \mod \mathbb{Z}$, introduced in the previous section. In contrast, if a single, uncoupled mode is populated, for instance the mode $a_1$, then the evolution in time is given by
\begin{equation}
|a_1\left( t \right) \rangle =  \frac{ 1+\cos{\left( 2t \right)}}{2} |a_1\rangle + i \frac{ 1 - \cos{\left( 2t \right) }}{2}|b_1\rangle + \frac{\sin{\left( 2t\right)} }{2} \left( |a_2 \rangle + i |b_2\rangle\right)
\end{equation}
The expectation value of the position is given by $\langle a_1 \left( t \right)| \hat{m} |a_1 \left( t \right) \rangle = \frac{5-\cos{\left( 4t\right)}}{4}$, with an average value in time of $5/4$. This deviates from the protected value of $n+1/2 \mod \mathbb{Z}$ because the state is an arbitrary state.

For the particular parameters we have just studied, the edge modes are extremely localized at the left and right boundaries. As soon as we depart from the strong-coupling limit or from $\phi=\pi$, the edge modes extend into the bulk of the system and the modes start to hybridize. Because of this, in a single-plaquette system, the winding number that can be extracted from transport measurements reveals little, if any, topological information. In fact, in a single plaquette, the four eigenstates of the Hamiltonian have the same constant value of the polarization independent of $\phi$. That is, the special significance of $\phi=\pi$ is not revealed by measuring the polarization.

\subsection{Scattering Matrix}


Following Ref.~\cite{Lecocq17}, the scattering model used for the fitting can also be derived as the Heisenberg equations of motion of an effective non-Hermitian model Hamiltonian describing four parametrically coupled resonator modes with frequencies  $\omega_a$, $\omega_b$, $\omega_c$ and $\omega_d$, loss rates $\kappa_a$ to $\kappa_d$ , and a time-dependent
coupling rate per link $ g_{nm}(t)  = |g_{nm}|\cos{(\omega_{nm}^pt + \phi_{nm})}$. The Hamiltonian is
\begin{equation}
 \hat{\mathcal{H}}/\hbar =  \sum_{n=1}^{4} \left(\omega_{n} - i\frac{\kappa_n}{2}\right)\hat{a}_n^\dagger \hat{a}_n + i \sqrt{\kappa_n^{\rm ext}} (\hat{a}_{{\rm i}, n} - \hat{a}_{{\rm i}, n}^\dagger)(\hat{a}_n^\dagger + \hat{a}_n)
 - \sum_{m  \neq n} g_{nm}(t)(\hat{a}_n^\dagger + \hat{a}_n)(\hat{a}_m^\dagger + \hat{a}_m) ,
\end{equation}
where $\kappa_{n}^{\rm ext}$ is the coupling rate to the external port n, $a_n$ and $a_{ {\rm i},n}$ are the time-dependent annihilation operators for the internal mode and input modes for each, respectively.  The total loss rates $\kappa_n = \kappa_n^{\rm int} + \kappa_{n}^{\rm ext}$ are the sum of the internal loss rates $\kappa_n^{\rm int}$ and the external loss rates $\kappa_{n}^{\rm ext}$, which arise from the coupling to the external transmission line.

Assuming that the cavities are probed at $\omega_n^{\rm s}$ and $\omega_m^{\rm s}$ where $\omega_m^{\rm s} > \omega_n^{\rm s}$. The pump tone frequency is set at $\omega_{\rm p}^{\rm s} = \omega_m^{\rm s} - \omega_n^{\rm s}$. Moving to the interaction picture and taking the appropriate rotating-wave approximation, we find
\begin{equation}
 \hat{\mathcal{H}}_{int}/\hbar = \sum_{n=1}^{4} \left(\omega_n - \omega_n- i\frac{\kappa_n}{2}\right)\hat{a}_n^\dagger \hat{a}_n
  + i \sqrt{\kappa_n^{\rm ext}} (\hat{a}_{ {\rm i}, n} \hat{a}_{n}^\dagger  - \hat{a}_{ {\rm i}, n}^\dagger \hat{a}_{n} ) 
 - \sum_{m  \neq n}\frac{|g_{nm}|}{2}(e^{i\phi_{nm}} \hat{a}_n^\dagger \hat{a}_m  +e^{-i\phi_{nm}} \hat{a}_n \hat{a}_m^\dagger ) .
\end{equation}
The steady-state equations of motion of the steady-state solutions then follows as 
\begin{equation}
 \begin{split}
 i \sqrt{\kappa_n^{\rm ext}} \hat{a}_{ {\rm i}, n} =  \left(\omega_n^{\rm s} - \omega_n + i\frac{\kappa_n}{2}\right)\hat{a}_n +  \sum_{m  \neq n} \frac{|g_{nm}|}{2}e^{i\phi_{nm}} \hat{a}_m
  \end{split}
\end{equation}
Defining the normalized coupling as $\beta_{nm} = |g_{nm}|e^{i\phi_{nm}}/2\sqrt{\kappa_n\kappa_m}$, and the normalized detuning is defined as
 \begin{equation}
     \tilde{\Delta}_n = \frac{\omega_n^{\rm s} - \omega_n}{\kappa_n} + \frac{i}{2}
 \end{equation}
 The equation of motion becomes
\begin{equation}
 \begin{split}
 i \sqrt{\kappa_n^{\rm ext}} \hat{a}_{ {\rm i}, n} =  \kappa_n \tilde{\Delta}_n \hat{a}_n + \sum_{m  \neq n} \sqrt{\kappa_n \kappa_m} \beta_{nm} \hat{a}_m.
  \end{split}
\end{equation}

For larger systems, formulating the EOMs describing the system in matrix form becomes useful. First, we define a number of vectors representing the cavity mode operators as $\bold{A} = (\hat{a}_{1},\dots,\hat{a}_{n})^T$, the input modes $\bold{A}_{\rm in} = (\hat{a}_{ {\rm i}, 1},\dots, \hat{a}_{ {\rm i }, n})^T$  and output modes $\bold{A}_{\rm out} = (\hat{a}_{ {\rm o}, 1 }\dots, \hat{a}_{ {\rm o}, n})^T$.
Also, we define the total loss matrix  $\bold{K} = {\rm diag} (\sqrt{\kappa_{1}},\dots,\sqrt{\kappa_{n}})$
and the external loss $\bold{K^{\rm ext}} =  {\rm diag}(\sqrt{\kappa^{\rm ext}_{1}},\dots ,\sqrt{\kappa_{n}^{\rm ext}})$ and finally the coupling matrix:-
\begin{equation*}
\bold{M}=\begin{pmatrix}
\tilde{\Delta}_1 & \beta_{12}  & \dots  \\
 \vdots  & \ddots &   \\
 \beta^*_{m1} &   &\tilde{\Delta}_m
\end{pmatrix}.
\end{equation*}
The final matrix form of the EOMs becomes $\bold{KMKA} = i \bold{K^{\rm ext} A_{\rm in}} $. 

To solve for the scattering matrix, we use the input-output formalism to define the output mode operators as $\bold{A}_{\rm out} = \bold{K}^{\rm ext} \bold{A} - \bold{A}_{\rm in}$. Finally, the steady-state scattering matrix is defined as 
\begin{equation}~\label{eq:ScatMatrix}
\bold{S} = \braket{\bold{A}_{\rm out}}/\braket{\bold{A}_{\rm in}^T} = i  \bold{HM^{-1}H } - \mathbbm{1}   
\end{equation}
where we have introduced $\bold{H} = {\rm diag}( \eta_1,\dots,\eta_n)$.

As a simple example, let us assume that the system is in the strong-coupling limit of the Creutz ladder, i.e. $g_{nm}=g$ for horizontal and diagonal bonds, $g_{nm}=0$ for vertical bonds. Further, driving the system at resonance and with no internal losses we have
$\kappa_{m} = \kappa_m^{\rm ext} =\kappa$ and define $\beta=g/2\kappa$.  Then, the scattering matrix
\begin{equation*}
\mathbb{S}=\begin{pmatrix}
\frac{1}{1+16 \beta^2} & -1 + \frac{1}{1+16 \beta^2} & \frac{4i\beta}{1+16 \beta^2} & \frac{4i\beta}{1+16 \beta^2} \\
-1 + \frac{1}{1+16 \beta^2} & \frac{1}{1+16 \beta^2} & \frac{4i\beta}{1+16 \beta^2} & \frac{4i\beta}{1+16 \beta^2} \\
\frac{4i\beta}{1+16 \beta^2} &\frac{4i\beta}{1+16 \beta^2} &\frac{1}{1+16 \beta^2} & -1 + \frac{1}{1+16 \beta^2} \\
\frac{4i\beta}{1+16 \beta^2} &\frac{4i\beta}{1+16 \beta^2} & -1 + \frac{1}{1+16 \beta^2}&\frac{1}{1+16 \beta^2}
\end{pmatrix},
\end{equation*}
has two eigenmodes located at the two edges of the plaquette with eigenvalue equal to one, and two eigenmodes that extend over the whole plaquette with equal amplitude on every site. This general structure is what is observed for the twisted plaquette described in Fig.~2 and Fig.~3 of the main text.

\bibliography{supplement}